\documentclass[pre,showpacs,twocolumn,preprintnumbers,amsmath,amssymb]{revtex4}

\usepackage{graphicx}
\usepackage{dcolumn}
\usepackage{bm}
\newcommand{\comment}[1]{}

\begin{document}

\preprint{APS/123-QED}

\title{Simulations of Aging and Plastic Deformation in Polymer Glasses}

\author{Mya Warren}
\email{mya@phas.ubc.ca}
\author{J{\"o}rg Rottler}%

\affiliation{Department of Physics and Astronomy, The University of
British Columbia, 6224 Agricultural Road, Vancouver, BC, V6T 1Z1,
Canada}

\date{\today}

\begin{abstract}
We study the effect of physical aging on the mechanical properties of
a model polymer glass using molecular dynamics simulations. The creep
compliance is determined simultaneously with the structural relaxation
under a constant uniaxial load below yield at constant
temperature. The model successfully captures universal features found
experimentally in polymer glasses, including signatures of mechanical
rejuvenation. We analyze microscopic relaxation timescales and
show that they exhibit the same aging characteristics as the
macroscopic creep compliance. In addition, our model indicates that
the entire distribution of relaxation times scales identically with
age. Despite large changes in mobility, we observe comparatively
little structural change except for a weak logarithmic increase in the
degree of short-range order that may be correlated to an observed
decrease in aging with increasing load.
\end{abstract}

\maketitle

\section{Introduction}
\label{sec:introduction} 

Glassy materials are unable to reach equilibrium over typical
experimental timescales \cite{Leshouches,angell_Science,
angell_JAP}. Instead, the presence of disorder at temperatures below
the glass transition permits only a slow exploration of the
configurational degrees of freedom. The resulting structural
relaxation, also known as physical aging \cite{Struik}, is one of the
hallmarks of glassy dynamics and leads to material properties that
depend on the wait time $t_w$ since the glass was formed. While
thermodynamic variables such as energy and density typically evolve
only logarithmically, the relaxation times grow much more rapidly with
wait time \cite{Struik,Hutchinson,angell_JAP}.

Aging is a process observed in many different glassy systems,
including colloidal glasses \cite{abou}, microgel pastes
\cite{cloitre}, and spin glasses \cite{Jonsson_spin_glass}, but is
most frequently studied in polymers due to their good glass-forming
ability and ubiquitous use in structural applications. Of particular
interest is therefore to understand the effect of aging on their
mechanical response during plastic deformation \cite{Hutchinson}.  In
a classic series of experiments, Struik \cite{Struik} studied many
different polymer glasses and determined that their stiffness
universally increases with wait time. However, it has also been found
that large mechanical stimuli can alter the intrinsic aging dynamics
of a glass. Cases of both decreased aging (rejuvenation) \cite{Struik}
and increased aging (overaging) \cite{Lequeux_PRL89,Lacks_PRL93} have
been observed, but the interpretation of these findings in terms of
the structural evolution remains controversial
\cite{McKenna_JPhys15,Struik_POLY38}.

The formulation of a comprehensive molecular model of the
non-equilibrium dynamics of glasses has been impeded by the fact that
minimal structural change occurs during aging. Traditional
interpretations of aging presume that structural relaxation is
accompanied by a decrease in free volume available to molecules and an
associated reduction in molecular mobility \cite{Struik}. While this
idea is intuitive, it suffers from several limitations. First, the
free volume has been notoriously difficult to define
experimentally. Also, this model does not seem compatible with the
observed aging in glassy solids under constant volume conditions
\cite{Utz_PRL84}, and cannot predict the aging behavior under complex
thermo-mechanical histories. Modern energy landscape theories describe
the aging process as a series of hops between progressively deeper
traps in configuration space \cite{Bouchaud_JPA,Sollich_PRL78}. These
models have had some success in capturing experimental trends, but
have yet to directly establish a connection between macroscopic
material response and the underlying molecular level processes. Recent
efforts to formulate a molecular theory of aging are promising but
require knowledge of how local density fluctuations control the
relaxation times in the glass \cite{Schweizer}.

Molecular simulations using relatively simple models of glass forming
solids, such as the binary Lennard-Jones glass \cite{Kob_JPhysCM11} or
the bead spring model \cite{Bennemann} for polymers, have shown rich
aging phenomenology. For instance, calculations of particle
correlation functions have shown explicitly that the characteristic
time scale for particle relaxations increases with wait time
\cite{Barrat_PRL78}. Recent work \cite{Utz_PRL84,Barrat_JChemPhys120}
has focused on the effect of aging on the mechanical properties;
results showed that the shear yield stress (defined as the overshoot
or maximum of the stress-strain curve) in deformation at constant
strain rate generally increases logarithmically with $t_w$. Based on a
large number of simulations at different strain rates and
temperatures, a phenomenological rate-state model was developed that
describes the combined effect of rate and age on the shear yield
stress for many temperatures below the glass transition
\cite{Rottler_PRL95}.

In contrast to the strain-controlled studies described above,
experiments on aging typically impose a small, constant stress and
measure the resulting creep as a function of time and $t_w$
\cite{Struik}. In this study, we perform molecular dynamics
simulations on a coarse grained, glass forming polymer model in order
to investigate the relationship between macroscopic creep response and
microscopic structure and dynamics. In Section \ref{ssec:mechanical},
we determine creep compliance curves for different temperatures and
applied loads (in the sub-yield regime) and find that, as in
experiments, curves for different ages can be superimposed by
rescaling time. The associated shift factors exhibit a power-law
dependence on the wait time, and the effect of aging can be captured
by an effective time as originally envisioned by Struik
\cite{Struik}. In Section \ref{ssec:microscopic}, we compute
microscopic mobilities and the full spectrum of relaxation times and
show their relationship to the creep response. Additionally, we study
several parameters that are sensitive to the degree of short-range
order in Section \ref{ssec:structure}. We detect very little evolution
toward increased local order in our polymer model, indicating that
short range order is not a sensitive measure of the mechanical
relaxation times responsible for the creep compliance of glassy
polymers.

\section{Simulations}
\label{sec:simulation}

We perform molecular dynamics (MD) simulations with a well-known model
polymer glass on the bead-spring level.  The beads interact via a
non-specific van der Waals interaction given by a 6-12 Lennard-Jones
potential, and the covalent bonds are modeled with a stiff spring that
prevents chain crossing \cite{Kremer_Grest}. This level of modeling
does not include chemical specificity, but allows us to study longer
aging times than a fully atomistic model and seems appropriate to
examine a universal phenomenon found in all glassy polymers. All
results will be given in units of the diameter $a$ of the bead, the
mass $m$, and the Lennard-Jones energy scale, $u_0$. The
characteristic timescale is therefore $\tau_{LJ}=\sqrt{ma^2/u_0}$, and
the pressure and stress are in units of $u_0/a^3$.  The Lennard-Jones
interaction is truncated at $1.5a$ and adjusted vertically for
continuity. All polymers have a length of 100 beads, and unless
otherwise noted, we analyze 870 polymers in a periodic simulation
box. Results are obtained either with one large simulation containing
the full number of polymers, or with several smaller simulations, each
starting from a unique configuration, whose results are averaged. The
large simulations and the averaged small simulations provide identical
results. The small simulations are used to estimate uncertainties
caused by the finite size of the simulation volume.

To create the glass, we begin with a random distribution of chains and
relax in an ensemble at constant volume and at a melt temperature of
$1.2u_0/k_B$. Once the system is fully equilibrated, it is cooled over
$750\tau_{LJ}$ to a temperature below the glass transition at $T_g
\approx 0.35u_0/k_B$ \cite{Bennemann}. The density of the melt is
chosen such that after cooling the pressure is zero. We then switch to
an NPT ensemble - the pressure and temperature are controlled via a
Nos\'e-Hoover thermostat/barostat - with zero pressure and age for
various wait times ($t_w$) between 500 to 75,000$\tau_{LJ}$. The aged
samples undergo a computer creep experiment where a uniaxial tensile
stress (in the z-direction) is ramped up quickly over $75\tau_{LJ}$,
and then held constant at a value of $\sigma$, while the strain
$\epsilon={\Delta}L_z/L_z$ is monitored.  After an initial elastic
deformation, the glass slowly elongates in the direction of applied
stress due to structural relaxations. In the two directions
perpendicular to the applied stress, the pressure is maintained at
zero.

\section{Results}
\label{sec:results}

\subsection{Macroscopic Mechanical Deformation}
\label{ssec:mechanical}

\begin{figure}[tbp]  
\begin{center}
\includegraphics[width=8cm]{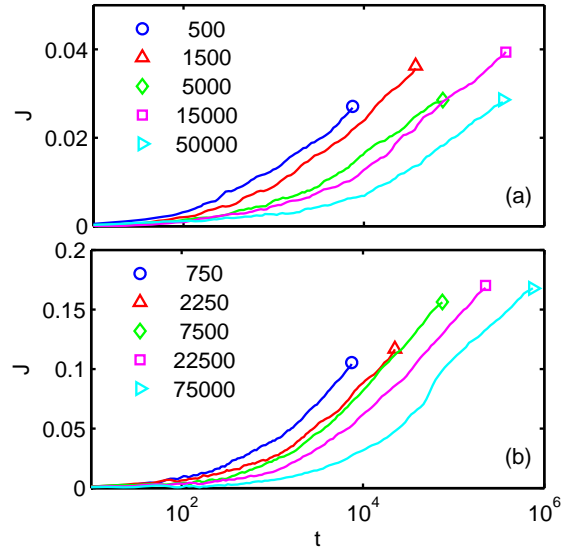}
\caption{Simulated creep compliance $J(t,t_w)$ at a glassy temperature
of $T=0.2u_0/k_B$ for various wait times $t_w$ (indicated in the
legend in units of $\tau_{LJ}$). A uniaxial load of (a)
$\sigma=0.4u_0/a^3$ and (b) $\sigma=0.5u_0/a^3$ is applied to the aged
glasses. The strain during creep is monitored over time to give the
creep compliance.}
\label{fig:creep}
\end{center} 
\end{figure}

\begin{figure}[tbp]  
\begin{center}
\includegraphics[width=8cm]{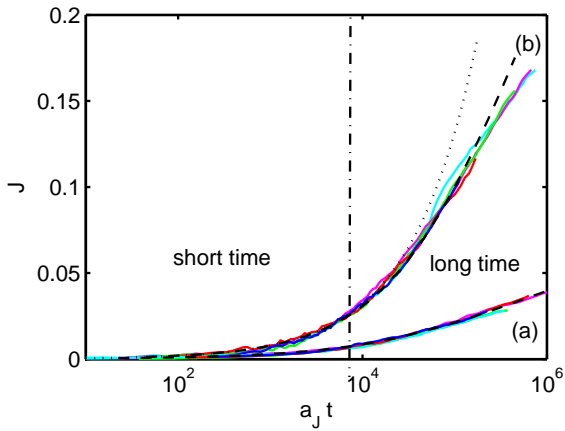}
\caption{The same data as Fig.~\ref{fig:creep} is shown with the
curves shifted by $a_J(t_w)$ to form a master curve. The dashed lines are
fits to the master curves using the effective time formulation, and
the dotted line is a short-time fit for comparison (see text).}
\label{fig:MC_fit_creep}
\end{center} 
\end{figure}

Historically, measurements of the creep compliance have been
instrumental in probing the relaxation dynamics of glasses, and
continue to be the preferred tool in investigating the aging of glassy
polymers \cite{Lequeux_JSM, Brinson_JSS32,Sollich_PRL78}. In his
seminal work on aging in polymer glasses, Struik \cite{Struik}
performed an exhaustive set of creep experiments on different
materials, varying the temperature and the applied load. In this
section, we perform a similar set of experiments with our model
polymer glass. The macroscopic creep compliance is defined as
\begin{equation}
J(t,t_w)=\frac{\epsilon(t,t_w)}{\sigma}.
\label{eqn:J}
\end{equation}
Compliance curves $J(t,t_w)$ for several temperatures and stresses
were obtained as a function of wait time since the quench;
representative data is shown in Figure \ref{fig:creep}. The curves for
different wait times appear similar and agree qualitatively with
experiment. An initially rapid rise in compliance crosses over into a
slower, logarithmic increase at long times. The crossover between the
two regimes increases with increasing wait time.  Struik showed that
experimental creep compliance curves for different ages can be
superimposed by rescaling the time variable by a shift factor, $a_J$,
\begin{equation}
J(t,t_w)=J(ta_J,t_w').
\label{eqn:master}
\end{equation}
This result is called time-aging time superposition
\cite{Struik,Hutchinson}. Simulated creep compliance curves from
Fig.~\ref{fig:creep} can similarly be superimposed, and the resulting
master curve is shown in Fig.~\ref{fig:MC_fit_creep}.

\begin{figure}[tbp]  
\begin{center}
\includegraphics[width=8cm]{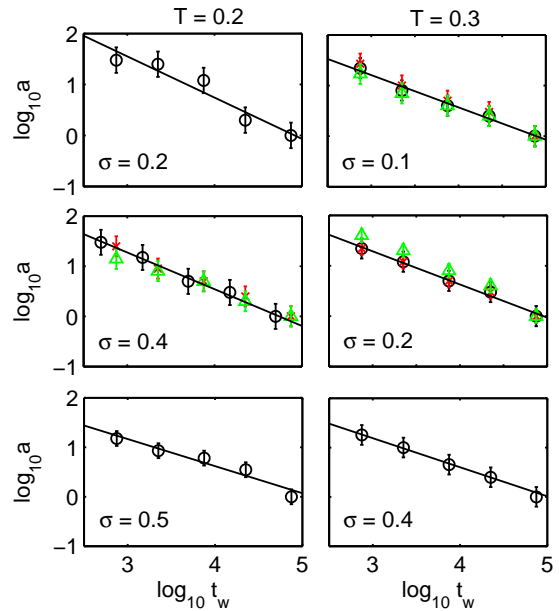}
\caption{Plot of the shift factors found by superimposing the creep
compliance curves, $a_J$ (circles), the mean-squared displacement
curves, $a_{MSD}$ (triangles), and the incoherent scattering function
curves, $a_C$ ($\times$) at different wait times (see text). The solid lines
are linear fits to the data.}
\label{fig:shift_creep}
\end{center} 
\end{figure}

Shift factors required for this data collapse are plotted versus the
wait time in Fig.~\ref{fig:shift_creep}. All data fall along a
straight line in the double-logarithmic plot, clearly indicating power
law behavior:
\begin{equation}
a_J{\sim}t_w^{-\mu}.
\label{eqn:aging}
\end{equation}
This power law in the shift factor is characteristic of aging.  $\mu$
is called the aging exponent, and has been found experimentally to be
close to unity for a wide variety of glasses in a temperature range
near $T_g$ \cite{Struik}.

\begin{figure}[tbp]  
\begin{center}
\includegraphics[width=8cm]{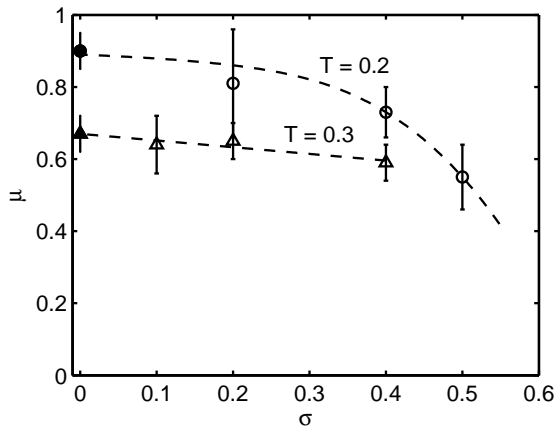}
\caption{The aging exponent, $\mu$, determined from the slopes of
$\log(a_J)$ versus $\log(t_w)$ (from Fig.~\ref{fig:shift_creep}) 
plotted versus stress (open symbols). The solid symbols at zero stress
refer to shift factors determined from $a_{MSD}$ (eq.~\ref{eqn:msd}) and
$a_C$ (eq.~\ref{eqn:Cq}) data only. The dashed lines are guides to the
eye.}
\label{fig:mu_vs_stress}
\end{center} 
\end{figure}

Figure \ref{fig:mu_vs_stress} shows the effect of stress and
temperature on the aging exponent, as determined from linear fits to
the data in Fig.~\ref{fig:shift_creep}. At $T=0.2u_0/k_B$, $\mu$ is
close to one for small stresses, but decreases strongly with
stress. This apparent erasure of aging by large mechanical
deformations has been called ``mechanical rejuvenation"
\cite{McKenna_PES}. Experiments have frequently found a stress
dependence of the aging exponent \cite{Struik}, although it is not
always the case that the aging process slows down with applied stress;
stress has been known to increase the rate of aging in some
circumstances as well \cite{Lequeux_PRL89,Lacks_PRL93}. The structural
origins of this effect are not well understood
\cite{McKenna_JPhys15,Struik_POLY38}.

At $T=0.3u_0/k_B$, we find that the aging exponent is somewhat smaller
than at $T=0.2u_0/k_B$ and varies much less with applied stress. This
behavior is most likely due to the fact that the temperature is
approaching $T_g$. Indeed, experiments show that $\mu$ rapidly drops
to zero above $T_g$. 
compliance is an order of magnitude larger at $T=0.3u_0/k_B$ than at
$T=0.2u_0/k_B$ and the data does not fully superimpose in a master
curve for long times where $J>0.2u_0/a^3$. Shift factors were obtained
from the small creep portion of the curves.

The relatively simple relationship between shift factors and wait time
permits construction of an expression that describes the entire master
curve in Fig.~\ref{fig:MC_fit_creep}.  For creep times that are short
compared to the wait time - such that minimal physical aging occurs
over the timescale of the experiment - experimental creep compliance
curves can be fit to a stretched exponential (typical of processes
with a spectrum of relaxation times),
\begin{equation}
J(t)=J_0\exp[(t/t_0)^m]
\label{eqn:stretched_exp}
\end{equation}
where $t_0$ is the retardation factor, and the exponent, $m$, has been
found to be close to $1/3$ for most glasses \cite{Struik}. A fit of
this expression to our simulated creep compliance curves is shown in
Fig.~\ref{fig:MC_fit_creep} (dotted line). This expression is clearly
only consistent with the data at times $t<t_w$.  At times much longer
than the wait time, the creep compliance varies more slowly due to the
stiffening caused by aging during the course of the experiment. Struik
suggested that eq.~(\ref{eqn:stretched_exp}) could be extended to the
long-time creep regime, where the experimental timescale may be longer
than the wait time, by introducing an effective time to account for
the slowdown in the relaxation timescales:
\begin{equation}
t_{\rm eff}=\int_0^t\left(\frac{t_w}{t_w+t'}\right)^\mu dt'
\label{eqn:t_eff}
\end{equation}

Upon replacing $t$ with $t_{\rm eff}$, eq.~(\ref{eqn:stretched_exp})
may be used to describe the entire creep curve. Creep compliance
curves from Fig.~\ref{fig:MC_fit_creep} can indeed be fit to this form
(dashed lines) for a known wait time, $t_w$, and aging exponent,
$\mu$, as obtained from the master curve. We find $m \approx
0.5\pm0.1$ for all stresses at $T=0.2u_0/k_B$, and a relatively broad
range of values for $J_0$ and $t_0$ are consistent with the data. For
the simple thermo-mechanical history prescribed by the creep
experiment, Struik's effective time formulation appears to work quite
well.

The present results parallel those of a recent simulation study of the
shear yield stress in glassy solids \cite{Rottler_PRL95}. In this
work, the glassy solid was deformed at constant strain rate, and two
different regimes of strong and weak rate dependence emerged depending
on the time to reach the yield point relative to the wait time. In
order to rationalize these results, a rate-state model was developed
that accounted for the internal evolution of the material with age
through a single state variable $\Theta(t)$. This formulation
successfully collapses yield stress data for different ages and strain
rates in a universal curve by adapting the evolution of the state
variable during the strain interval. We note here that this state
variable is closely related to Struik's effective time, as it tries to
subsume the modified aging dynamics during deformation in a single
variable and in particular easily includes the effects of overaging or
rejuvenation.

\subsection{Microscopic Dynamics}
\label{ssec:microscopic}

\begin{figure}[tbp]  
\begin{center}
\includegraphics[width=8cm]{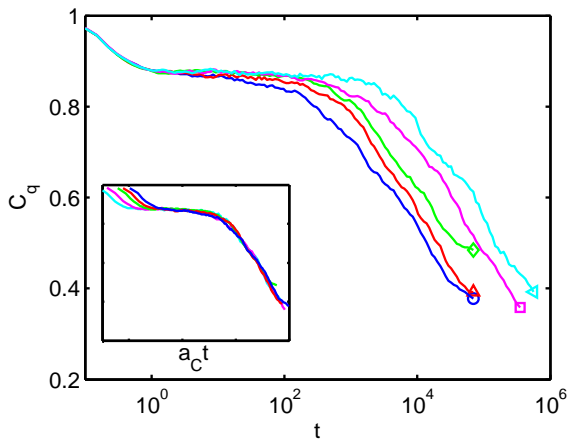}
\caption{Incoherent scattering factor (eq.~\ref{eqn:Cq}) for different
wait times measured under the same loading conditions as in
Fig.~\ref{fig:creep}(a) for $q=(0,0,2\pi)$. The inset shows the master
curve created by rescaling the time variable by $a_C$. Symbols as in
Fig.~\ref{fig:creep}(b).}
\label{fig:Cq}
\end{center} 
\end{figure}

The aging behavior of the simulated mechanical response functions
agrees remarkably well with experiment. Additional microscopic
information from simulations allows us to obtain more directly the
relevant timescales of the system, and the relevant microscopic
processes responsible for aging. One parameter which has been useful
in studying glassy dynamics is the ``self" part of the incoherent
scattering factor \cite{Barrat_PRL78},
\begin{equation}
C_q(t,t_w)=\frac{1}{N}\sum^{N}_{j=1}exp(i\vec{q}\cdot[\vec{r}_j(t_w+t)-\vec{r}_j(t_w)])
\label{eqn:Cq}
\end{equation}
where $\vec{r_j}$ is the position of the $j^{th}$ atom, and $\vec{q}$
is the wave-vector.  $C_q$ curves as a function of age are shown in
Fig.~\ref{fig:Cq} and exhibit three distinct regions. Initially, $C_q$
decreases as particles make very small unconstrained excursions about
their positions.  There follows a long plateau, where the correlation
function does not change considerably. In this regime, atoms are not
free to diffuse, but are trapped in local cages formed by their
nearest neighbours. For this reason, the time spent in the plateau
regime is often associated with a ``cage time". The plateau region
ends when particles finally escape from local cages
($\alpha$-relaxation), and larger atomic rearrangements begin to take
place. The cage time corresponds closely to the transition from
short-time to long-time regime observed in the creep
compliance. Structural rearrangements taking place in the
$\alpha$-relaxation regime are clearly associated with the continued
aging observed in the creep compliance, as well as plastic
deformations occurring in that region.

The correlation functions for different ages are similar in form, but
the time spent in the plateau region increases with age. Just as creep
compliance curves can be shifted in time to form a master curve, we
may overlap the long-time, cage-escape regions of $C_q$ by rescaling
the time variable of the correlation data at different ages (see inset
of Fig.~\ref{fig:Cq}). The corresponding shift factors $a_C(t_w)$ are
also shown in Fig.~\ref{fig:shift_creep}, where we see that the
increase in cage time with age follows the same power law as the shift
factors of the creep compliance. These results are qualitatively
similar to the scaling of the relaxation times with age found in
\cite{Barrat_PRL78} with no load.

\begin{figure}[tbp]  
\begin{center}
\includegraphics[width=8cm]{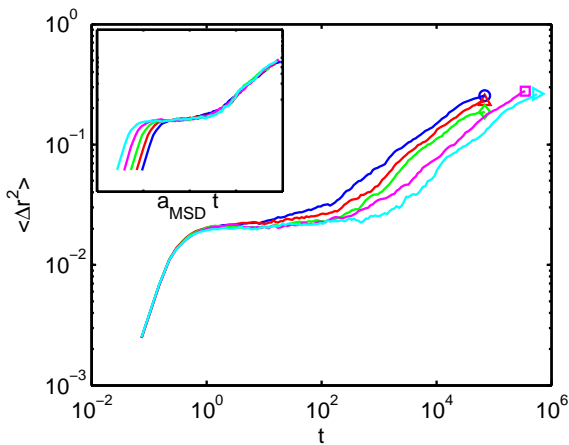}
\caption{Mean-squared displacement (eq.~\ref{eqn:msd}) for different
wait times measured under the same loading conditions as in
Fig.~\ref{fig:creep}(a). The inset shows the master curve created by
rescaling the time variable by $a_{MSD}$. Symbols as in
Fig.~\ref{fig:creep}(b).}
\label{fig:msd}
\end{center} 
\end{figure}

The real space quantity corresponding to $C_q$ is the mean squared
displacement,
\begin{equation}
\langle{\vec r}(t,t_w)^2\rangle=\frac{1}{N}\sum^{N}_{j=1} \Delta \vec{r}_j(t,t_w)^2
\label{eqn:msd}
\end{equation}
where $\Delta \vec{r}_j(t,t_w)=\vec{r}_j(t_w+t)-\vec{r}_j(t_w)$. This
function is shown in Fig.~\ref{fig:msd}. Again we see three
characteristic regions of unconstrained (ballistic), caged, and
cage-escape behavior. The departure from the cage plateau likewise
increases with age, and a master curve can be constructed by shifting
the curves with a factor $a_{MSD}$ (see inset of Fig.~\ref{fig:msd}).
Shift factors $a_{MSD}$ are plotted in Fig.~\ref{fig:shift_creep},
along with shifts for creep compliance and incoherent scattering
function. As anticipated, the shifts versus wait time for
$\left\langle {\Delta}r^2 \right\rangle$ fully agree with those
obtained from $C_q$ and $J$. This clearly demonstrates that for our
model, the cage escape time is indeed the controlling factor in the
aging dynamics of the mechanical response functions.

Additional information about microscopic processes can be obtained by
studying not only the mean of the displacements, but also the full
spectrum of relaxation dynamics as a function of time and wait
time. To this end, we measure the probability distribution $P(\Delta
r(t,t_w)^2)$ of atomic displacements during time intervals, $t$, for
glasses at various ages, $t_w$. This quantity is complementary to the
measurements of dynamical heterogeneities detailed in
\cite{Vollmayr_PRE72}, where the spatial variations of the vibrational
amplitudes were measured at a snapshot in time to show the
correlations of mobile particles in space. In our study, we omit the
spatial information, but retain all of the dynamical information.

\begin{figure}[tbp]  
\begin{center}
\includegraphics[width=8cm]{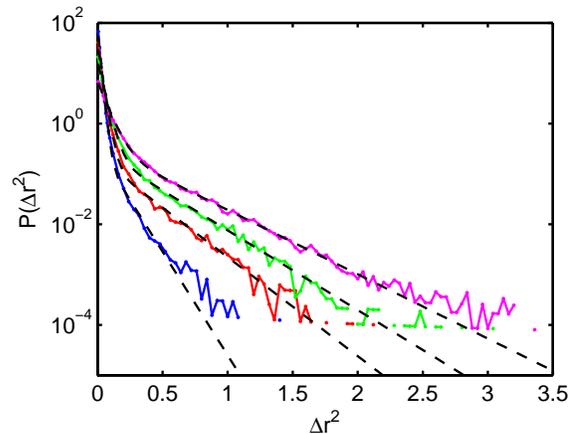}
\caption{The displacement probability distribution versus time
measured under the same loading conditions as in
Fig.~\ref{fig:creep}(a), with a wait time of $500\tau_{LJ}$. The solid
lines from left to right are obtained at times $t$ of $75$, $750$,
$7500$, and $75000\tau_{LJ}$. The dashed lines show fits to the double
Gaussian distribution (see text, eq.~\ref{eqn:msd_dist}).}
\label{fig:dist_vs_t}
\end{center} 
\end{figure}

Representative distribution functions are shown in
Fig.~\ref{fig:dist_vs_t} for a constant wait time of
$t_w=500\tau_{LJ}$ and various time intervals $t$.  The distributions
were obtained from a smaller system of only 271 polymer chains due to
memory constraints. The data does not reflect a simple Gaussian
distribution, but clearly exhibits the presence of two distinct
scales: there is a narrow distribution of caged particles and a wider
distribution of particles that have escaped from their cages. This
behavior can be described by the sum of two Gaussian peaks,
\begin{equation}
P(\Delta{r}^2)=N_1\exp\left(\frac{-\Delta{r}^2}{\sigma_1^2}\right)+N_2\exp\left(\frac{-\Delta{r}^2}{\sigma_2^2}\right).
\label{eqn:msd_dist}
\end{equation}
Deviations from purely Gaussian behavior are common in glassy systems
and are a signature of dynamical heterogeneities \cite{Kob_PRL79,Vollmayr_PRE72}. Experiments on colloidal glasses \cite{Weitz_CP284} show a
similar separation of displacement distributions into fast and slow
particles. 

\begin{figure}[tbp]  
\begin{center}
\includegraphics[width=8cm]{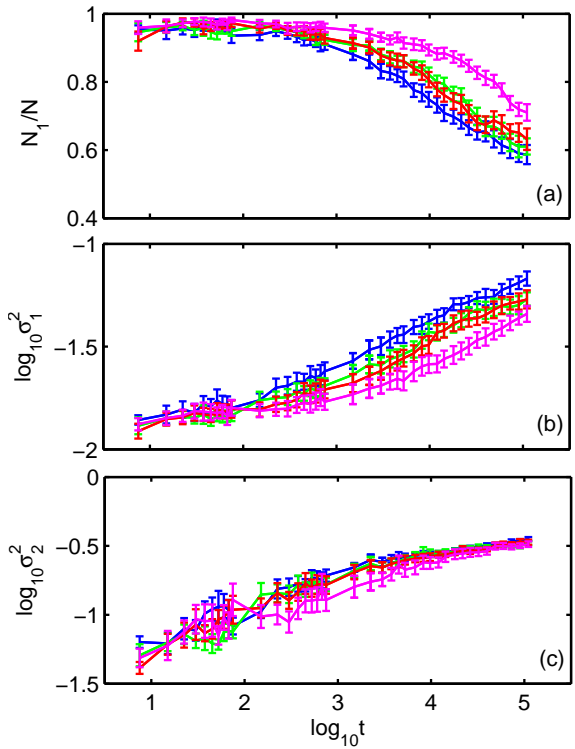}
\caption{The Gaussian fit parameters for the distribution of
displacements (see text, eq.~\ref{eqn:msd_dist}), (a) $N_1/N$, (b)
$\sigma_1^2$, and (c) $\sigma_2^2$ measured under the same loading
conditions as in Fig.~\ref{fig:creep}(a). The curves are for
wait times increasing from left to right from $500\tau_{LJ}$ to
$15000\tau_{LJ}$.}
\label{fig:fit_params_all}
\end{center} 
\end{figure}

A fit of the normalized distributions to eq.~(\ref{eqn:msd_dist})
(dashed lines in Fig.~\ref{fig:dist_vs_t}) requires adjustment of
three parameters: the variance of caged and mobile particle
distributions, $\sigma_1^2$ and $\sigma_2^2$, as well as their
relative contributions $N_1/N$, where $N=N_1+N_2$.  These parameters
are sufficient to describe the full evolution of the displacement
distribution during aging. In Fig.~\ref{fig:fit_params_all}, we show
the fit parameters as a function of time and wait time. Again two
distinct time scales are evident. At short times, most of the
particles are caged $(N_1/N\approx 1)$, and the variance of the cage
peak is also changing very little. There are few rearrangements in
this regime, however Fig.~\ref{fig:fit_params_all}(c) shows that a
small fraction of particles are mobile at even the shortest times. At
a time corresponding to the onset of cage escape, the number of
particles in the cage peak begins to rapidly decay, and the variance
of the cage peak increases.  This indicates that the cage has become
more malleable - small, persistent rearrangements occur leading to
eventual cage escape. In this regime, the variance of the mobile peak
increases very little. Note that the typical length scale of
rearrangements is less than a particle diameter even in the cage
escape regime, but the number of particles undergoing rearrangements
changes by more than 50\%.

\begin{figure}[tbp]  
\begin{center}
\includegraphics[width=8cm]{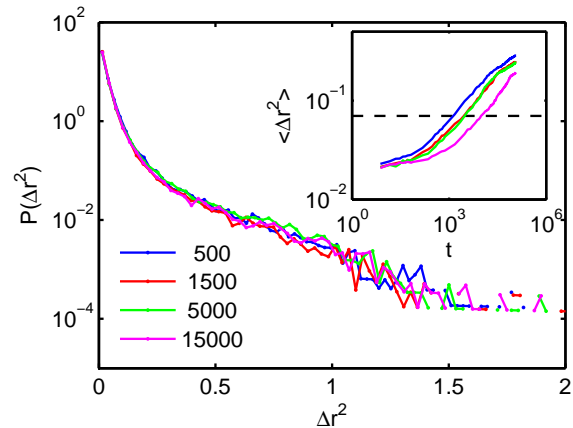}
\caption{The displacement probability distribution measured under the
same loading conditions and wait times as in Fig.~\ref{fig:creep}(a)
plotted at times corresponding to $<r^2(t,t_w)>=0.7$, shown in the
inset as a dashed line. The legend indicates the wait time.}
\label{fig:dist_scaled}
\end{center} 
\end{figure}

Similar to the compliance and mean-squared displacement curves, the
data in Fig.~\ref{fig:fit_params_all}(a) and (b) can also be
superimposed by shifting time.
Shift factors for $N_1/N$ and $\sigma_1^2$ coincide exactly with
shifts for the mean; however, data for $\sigma_2^2$
(Fig.~\ref{fig:fit_params_all}(c)) seems to be much less affected by
the wait time. The aging dynamics appears to be entirely determined by
the cage dynamics, and not by larger rearrangements within the glass.

Since the fit parameters exhibit the same scaling with wait time as
the mean, one might expect that the entire distribution of
displacements under finite load scales with the evolution of the mean.
In Fig.~\ref{fig:dist_scaled}, we plot displacement distributions for
several wait times at time intervals chosen such that the mean squared
displacements are identical (see inset). Indeed, we find that all
curves overlap, indicating that the entire relaxation spectrum ages in
the same way. A similar observation was recently made in simulations
of a model for a metallic glass aging at zero stress
\cite{Castillo_NaturePhys3}, although in this study the tails of the
distribution were better described by stretched exponentials.

\begin{figure}[tbp]  
\begin{center}
\includegraphics[width=8cm]{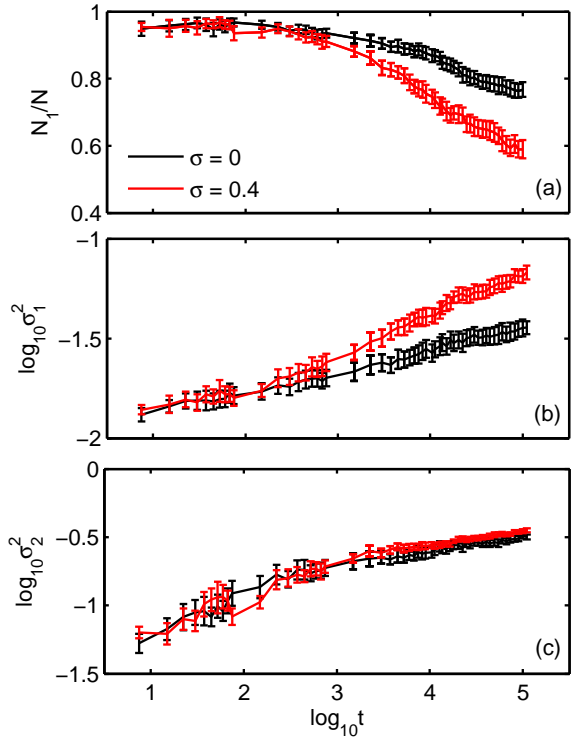}
\caption{The Gaussian fit parameters to the displacement distributions
(see text, eq.~\ref{eqn:msd_dist}) (a) $N_1/N$, (b)
$\sigma_1^2$, and (c) $\sigma_2^2$ of a sample aged at $T=0.2u_0/k_B$
for $t_w=500\tau_{LJ}$, and then either undergoing a creep experiment at
$\sigma=0.4u_0/a^3$ (black), or simply aging further at zero stress (red).}
\label{fig:fit_params_100000}
\end{center} 
\end{figure}

In order to study the effect of load on the relaxation dynamics, we
compare in Figure \ref{fig:fit_params_100000} the fit parameters for a
sample undergoing creep (replotted from Fig.~\ref{fig:fit_params_all})
and a reference sample without load. It is clear that the dynamics are
strongly affected by the applied stress only in the region
characterized by $\alpha$-relaxations. For the stress applied here,
the onset of cage-escape does not appear to be greatly modified by the
stress, however the decay in $N_1/N$ and the widening of the cage peak
are accelerated. The stress does not modify the variance of the mobile
peak, confirming again the importance of local rearrangements as
compared to large-scale motion in the dynamics of the system. The
accelerated structural rearrangements caused by the stress result in
creep on the macroscopic scale, but may also be responsible for the
modification of the aging dynamics with stress as observed in
Fig.~\ref{fig:mu_vs_stress}.

\subsection{Structural evolution}
\label{ssec:structure}

The connection between the dynamics and the structure of a glass
during aging remains uncertain, mostly because no structural parameter
has been found that strongly depends on wait time. Recent simulation
studies of metallic glasses have shown the existence of several short
range order parameters that can distinguish between glassy states
created through different quenching protocols
\cite{Falk_PRL95,Falk_JChemPhys122,Falk_PRB73}. A strong correlation
has been found between ``ordered'' regions of the glass and strain
localization. Many metallic glasses are known to form
quasi-crystalline structures that optimize local packing. It remains
to be seen whether the short-range order evolves in the context of
aging and in other glass formers such as polymers and colloids. A
recent experimental study of aging in colloidal glasses found no
change in the distribution function of a tetrahedral order parameter
\cite{Cianci}. Below, we investigate several measures of local order
in our model as they evolve with age and under load.

\begin{figure}[htbp]  
\begin{center}
\includegraphics[width=8cm]{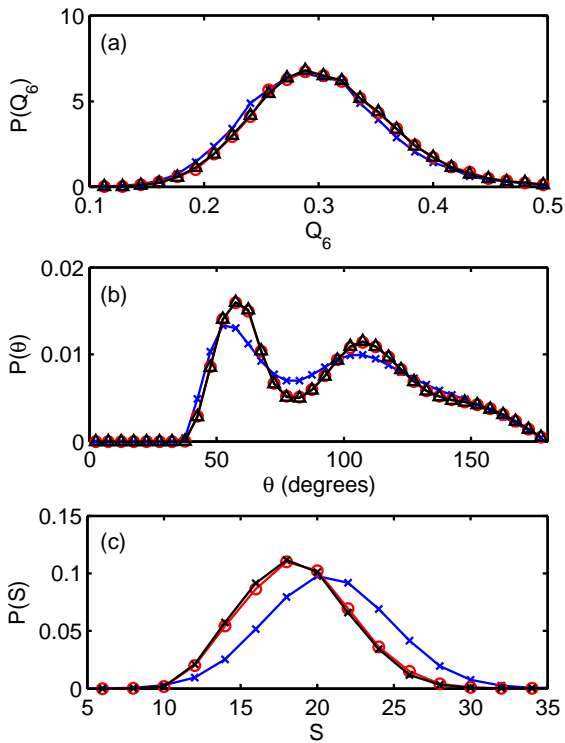}
\caption{Short-range order parameters: (a) the bond-orientational
parameter, (b) the three-body angular correlations, (c) the surface
triangulated order (see text for discussion). x's show the melt state,
circles show the sample aged for $t_w=500\tau_{LJ}$, and triangles
show the sample aged for $500,000\tau_{LJ}$ .}
\label{fig:order-all}
\end{center} 
\end{figure}

Since Lennard-Jones liquids are known to condense into a crystal with
fcc symmetry at low temperatures, it is reasonable to look for the
degree of local fcc order in our polymer model. The level of fcc order
can be quantified via the bond orientational parameter
\cite{Torquato_PRL84},
\begin{equation}
Q_6=\left(\frac{4\pi}{13}\sum^{6}_{m=-6}\left|\overline{Y_{6m}}\right|^2\right)^{1/2}.
\label{eqn:Q6}
\end{equation}
 This parameter has been successfully used to characterize the degree
of order in systems of hard sphere glasses. $Q_6$ is determined for
each atom by projecting the bond angles of the nearest neighbours onto
the spherical harmonics, $Y_{6m}(\theta,\phi)$. The overbar denotes an
average over all bonds. Nearest neighbours are defined as all
atoms within a cutoff radius, $r_c$, of the central atom. For all of
the order parameters discussed here, the cutoff radius is defined by
the first minimum in the pair correlation function, in this case
$1.45a$. $Q_6$ is approximately $0.575$ for a perfect fcc crystal; for
jammed structures, it can exhibit a large range of values less than
about 0.37 \cite{Torquato_PRL84}. The full distribution of $Q_6$ for
our model glass is shown for several ages as well as an initial melt
state in Fig.~\ref{fig:order-all}(a). We see that there is very little
difference even between melt and glassy states in our model, and
no discernible difference at all with increasing age.

Locally, close-packing is achieved by tetrahedral ordering and not fcc
ordering, however, tetrahedral orderings cannot span the system. The
glass formation process has been described in terms of frustration
between optimal local and global close-packing structures. To
investigate the type of local ordering in this model, we investigate a
3-body angular correlation function, $P(\theta)$. $\theta$ is defined
as the internal angle created by a central atom and individual pairs
of nearest-neighbours, and $P(\theta)$ is the probability of
occurrence of $\theta$. Results for this correlation are shown in
Fig.~\ref{fig:order-all}(b). Two peaks at approximately $60^\circ$ and
$110^\circ$ indicate tetrahedral ordering. The peaks sharpen under
quenching from the melt, but the distribution does not evolve
significantly during aging. In contrast, simulations of metallic glass
formers showed a stronger sensitivity of this parameter to the quench
protocol \cite{Falk_JChemPhys122}, but most of those changes may be
due to rearrangements in the supercooled liquid state and not in the
glassy state.

\begin{figure}[htbp]  
\begin{center}
\includegraphics[width=8cm]{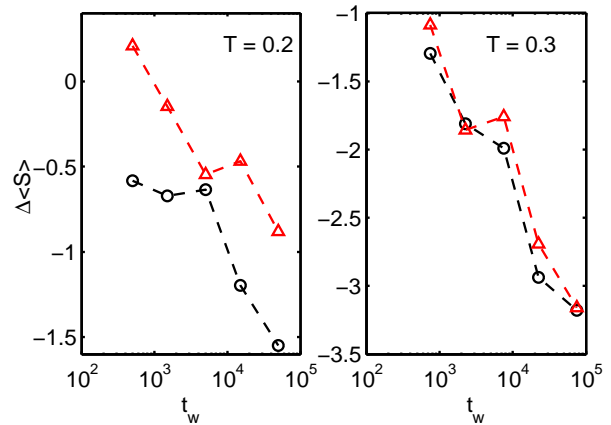}
\caption{Precent change in the triangulated surface order parameter
with wait time as compared to the just-quenched sample. Circles are
for samples aged at zero pressure for the time $t_w$. Triangles are
for the same samples immediately after ramping up to the creep
stress. For $T=0.2u_0/k_B$ this stress is $0.4u_0/a^3$, and for
$T=0.3u_0/k_B$ the stress is $0.1u_0/a^3$.}
\label{fig:sc_change}
\end{center} 
\end{figure}

Another parameter that has been successful in classifying glasses is
the triangulated surface order parameter \cite{Falk_PRB73},
\begin{equation}
S=\sum_q(6-q)\nu_q
\label{fig:SO}
\end{equation}
which measures the degree of quasi-crystalline order.  The surface
coordination number, $q$, is defined for each atom of the coordination
shell as the number of neighbouring atoms also residing in the
coordination shell; $\nu_q$ is the number of atoms in the coordination
shell with surface coordination $q$. Ordered systems have been
identified with $S$ equal to 12 (icosahedron), 14, 15 and 16. Figure
\ref{fig:order-all}(c) shows the probability distribution for $P(S)$
for the melt and for glassy states with short and long wait times. The
peak of the distribution moves toward lower $S$ (more ordered) upon
cooling, and continues to evolve slowly in the glass. The mean of $S$
relative to the as-quenched state, $\langle S\rangle$, is shown in
Fig.~\ref{fig:sc_change} as a function of wait time at two
temperatures.  We see that $\langle S\rangle$ is a logarithmically
decreasing function of wait time. Even though this is not a strong
dependence, this order parameter is significantly more sensitive to
age than the others that have been investigated.

Figure \ref{fig:sc_change} also shows the order parameter $\langle
S\rangle$ after the ramped-up stress has been applied to the aged
samples. We can see that at $T=0.2u_0/k_B$, some of the order that
developed during age is erased, while no appreciable change occurs at
the higher temperature $T=0.3u_0/k_B$. These results correlate well
with the behavior of the aging exponent found in
Fig.~\ref{fig:mu_vs_stress}, where mechanical rejuvenation was found
at lower temperatures but was much less pronounced at higher $T$. The
load is applied very quickly, and most of the deformation in this
regime is affine, however, the strain during this time was similar for
both temperatures, therefore the effect is not simply due to a
change in density. More work is needed to clarify the nature of the
structural changes during rejuvenation.

\section{Conclusions}
\label{sec:conclusions}

We investigate the effects of aging on both macroscopic creep response
and underlying microscopic structure and dynamics through simulations
on a simple model polymer glass. The model qualitatively reproduces
key experimental trends in the mechanical behavior of glasses under
sustained stress. We observe a power-law dependence of the relaxation
time on the wait time with an aging exponent of approximately unity,
and a decrease in the aging exponent with increasing load that
indicates the presence of mechanical rejuvenation. The model creep
compliance curves can be fit in the short and long-time regimes using
Struik's effective time formulation. Additionally, investigation of
the microscopic dynamics through two-time correlation functions has
shown that, for our model glass, the aging dynamics of the creep
compliance exactly corresponds to the increase in the cage escape or
$\alpha$-relaxation time.

A detailed study of the entire distribution of particle displacements
yields an interesting picture of the microscopic dynamics during
aging. The distribution can be described by the sum of two Gaussians,
reflecting the presence of caged and mobile particles. The fraction of
mobile particles and the amplitude of rearrangements in the cage
strongly increase at the cage escape time. However, in analogy with
results in colloidal glasses \cite{Weeks_JCM15}, structural
rearrangements occur even for times well within the ``caged'' regime,
and fairly independent of wait time and stress.
For our model glass, we find that the entire distribution of
displacements scales with age in the same way as the mean. At times
when the long-time portion of the mean squared displacement overlaps,
the distribution of displacements at different wait times completely
superimpose, confirming that all of the mechanical relaxation times
scale in the same way with age.

To characterize the evolution of the structure during aging, we
investigate several measures of short-range order in our model
glass. We find that the short-range order does not evolve strongly
during aging. The triangulated surface order \cite{Falk_PRB73},
however, shows a weak logarithmic dependence on age. Results also show
a change in structure when a load is rapidly applied, and this seems
to be correlated with an observed decrease in the aging exponent under
stress.

This study has characterized the dynamics of a model glass prepared by
a rapid quench below $T_g$, followed by aging at constant $T$ and
subsequent application of a constant load. For such simple
thermo-mechanical histories, existing phenomenological models work
well, however, the dynamics of glasses are in general much more
complex. For instance, large stresses in the non-linear regime modify
the aging dynamics and cause nontrivial effects such as mechanical
rejuvenation and over-aging \cite{McKenna_JPhys15,Lacks_PRL93}. Also,
experiments have shown that the time-age time superposition no longer
holds when polymer glasses undergo more complex thermal histories such
as a temperature jump \cite{Lequeux_JSM}. The success of our study in
analyzing simple aging situations indicates that the present
simulation methodology will be able to shed more light on these
topics in the near future.

\begin{acknowledgments}
We thank the Natural Sciences and Engineering Council of Canada
(NSERC) for financial support. Computing time was provided by
WestGrid. Simulations were performed with the LAMMPS molecular
dynamics package \cite{LAMMPS}.
\end{acknowledgments}


\end{document}